\documentclass[11pt,a4paper]{article}
\newcommand{\be}{\begin{equation}}
\newcommand{\ee}{\end{equation}}
\newcommand{\bi}{\begin{itemize}}
\newcommand{\ei}{\end{itemize}}
\newcommand{\bea}{\begin{eqnarray}}
\newcommand{\eea}{\end{eqnarray}}
\newcommand{\ba}{\begin{array}}
\newcommand{\ea}{\end{array}}

\usepackage[dvipsnames]{xcolor}

\usepackage[left=2.0cm, right=2.0cm, top=2.50cm, bottom=2.50cm]{geometry}
\usepackage[utf8]{inputenc}
\usepackage{cancel}
\usepackage[english]{babel}
\usepackage{physics}
\usepackage{amsthm}
\usepackage{mathrsfs}
\usepackage{mathtools}
\usepackage{graphicx}
\usepackage{changepage}
\usepackage{amssymb,amsmath}
\usepackage{verbatim}
\usepackage{empheq}
\usepackage{xcolor}
\usepackage{tikz}
\usepackage{float}
\usepackage{multirow}
\usepackage{multicol}
\usepackage{centernot}
\usepackage{csquotes}
\usepackage[colorlinks=true, allcolors=blue]{hyperref}

\usepackage{subfig}

\usepackage[hang,flushmargin]{footmisc} 
\usepackage{caption}
\usepackage[normalem]{ulem}

\usepackage[
    bibstyle=chem-rsc, 
    citestyle=numeric-comp, 
    backend=biber, 
    sorting=none, 
    maxnames=99,
    url=true,
]{biblatex}
\usepackage{booktabs}
\usepackage{siunitx}
\usepackage{longtable}

\hypersetup{final}

\AtBeginDocument{\RenewCommandCopy\qty\SI}

\title{Geometry-controlled competition between axis centering and detwinning in fivefold-twinned gold nanoparticles}
\author{Silvia Fasce, Diana Nelli, Luca Benzi, Georg Daniel Förster,
Riccardo Ferrando}

\begin{document}

\bigskip
\Large
\noindent

{\bf Geometry-controlled competition between axis centering and \\ detwinning in fivefold-twinned gold nanoparticles}

\par
\rm
\normalsize
\hrule

\vspace{.3cm}

\noindent
{Silvia Fasce$^{1}$}, {Diana Nelli$^{1,*}$},{ Luca Benzi$^{1,2}$}, {Georg Daniel Förster $^2$}, {Riccardo Ferrando $^1$}\\

\par
\small
\noindent$^1$ Physics Department, University of Genoa, Genoa, Italy\\
\smallskip
\noindent$^2$ ICMN, CNRS, Université d'Orléans, Orléans, France\\
\smallskip
\noindent$^*$ \underline{diana.nelli@edu.unige.it}
\normalsize

\begin{abstract}
\noindent 
Fivefold-twinned metal nanoparticles host a central wedge disclination that strongly influences their mechanical and catalytic properties. Yet the atomistic mechanisms governing the stability, migration, and annihilation of this topological defect remain incompletely understood. Here we present a systematic molecular dynamics study of gold Marks decahedra in which the fivefold axis is artificially brought close to the surface by controlled geometric modifications. By generating concave and convex morphologies with varying axis depth, we uncover a geometry-controlled competition between axis centering and detwinning. Concave geometries promote surface diffusion that restores fivefold symmetry, either by recentering the original disclination or by nucleating a new subsurface axis through collective atomic rearrangements. In contrast, convex structures with a shallow axis undergo rapid detwinning within the first nanoseconds via surface glide, leading to single-twin or fully FCC configurations. Remarkably, positioning the axis just two atomic layers beneath the surface suppresses detwinning and restores stability. Our results demonstrate that surface curvature and defect depth critically regulate disclination mobility and twin stability, providing a mechanistic framework to understand the structural evolution of multi-twinned nanoparticles and to guide the controlled design of defect-engineered nanomaterials.
\end{abstract}

\section{Introduction}

Crystallization processes at the nanoscale often lead to the formation of structures containing topological defects; among these, multi-twinned nanoparticles such as icosahedra (Ih) and decahedra (Dh) are of particular interest due to their unique non-crystalline motifs \cite{Baletto2005_aps,martin1996shells,xia2009shape,koga2003population,johnston2002atomic,marks1979multiply, Li2008Nature}. A decahedron consists of five FCC tetrahedra meeting at a central fivefold axis and separated by twin planes. The five tetrahedral subunits create an intrinsic deficit of $7.35^\circ$, because the fivefold symmetry is incompatible with the long-range translational order of the FCC lattice  \cite{Cleveland1997,Marks1994}. This ``angular deficiency" is the defining characteristic of a wedge disclination, a linear topological defect that carries substantial strain \cite{deWit1973}. For sufficiently small nanoparticles, the high surface-to-volume ratio causes surface-energy contributions to outweigh the strain cost, so the twinned decahedral morphology remains energetically favorable \cite{Ino1969_jpsj,Baletto2005_aps}.
Twin planes and dislocations have been shown to enhance numerous physical and chemical properties including mechanical strength, ductility, fatigue resistance, and electrical conductivity, across a range of materials \cite{li2010Nature,warner2007NatureMat,chen2003Science,Lu2009Science_rev,schiotz2003Science,Lu2009Science_stren,Huang2014Nature,Wei2014NatureComm,Pan2017Nature,Shute2011ActaM,Li2014NatComm,lu2004Scienceultrahigh,Anderoglu2008APL}.
Furthermore, they have been identified as critical sites that enhance the catalytic activity, selectivity, and electrochemical stability of nanoparticles \cite{bhuyan_perspective_2025, muhammad_defect_2024}. In particular, fivefold twin gold nanoparticles exhibit distinctive properties, including modified dipolar plasmonic resonance, superior mechanical resistance, and enhanced catalytic activity, all of which are intrinsically linked to crystallographic anisotropy, high defect density, and elevated internal strain \cite{zhou_decahedral_2019,Qi2008NatureMat, geng_grain_2024}.

Understanding the evolution of fivefold axes and twin planes, and exploiting these mechanisms to obtain desired fivefold-twinned configurations, is therefore crucial for tailoring the properties of catalysts and other advanced nanomaterials. 
However, the atomistic mechanisms driving structural evolution of multi-twinned nanoparticles, such as twin formation (twinning) and twin elimination (detwinning), are still not fully understood. Advanced in situ transmission electron microscopy (TEM), combined with complementary simulations, has provided direct observations of transformation and growth pathways \cite{qu_recent_2023, chao_situ_2023, Ma2020_acs,Ma2025_acs,Elkoraychy2022_trsc,chen_penta-twin_2021}, revealing collective twin‑boundary migration, the emergence of complex multi‑twinned morphologies, and rearrangements driven by particle attachment, external stress, or thermal annealing \cite{song_oriented_2020, wang_atomic-scale_2025, zhang_situ_2025, zhang_swap_2022, li_five-fold_2024, song_uneven_2024}. Examples include the formation of fivefold-twinned configurations via decomposition of high‑energy grain boundaries during oriented attachment  \cite{song_oriented_2020}, and detwinning mediated by dislocation slip and twin‑boundary migration under uneven strain distributions \cite{song_uneven_2024}.  However, capturing the rapid atomic dynamics governing these processes on nanosecond–to-microsecond timescales remains challenging. 
A promising approach to overcoming these experimental limitations is the use of atomistic simulations \cite{Elkoraychy2022_trsc,Nelli2020nanoscale,Settem2024am}.

Here, we present a systematic computational study of the mechanisms driving twinning and detwinning in gold nanoparticles, through the formation, evolution and annihilation of a surface-proximal fivefold axis. Our main objective is to investigate how geometry
influences structural evolution by comparing the behavior of concave and convex structures. 

\begin{figure}[h!]
    \centering
    \includegraphics[width=0.95\textwidth]{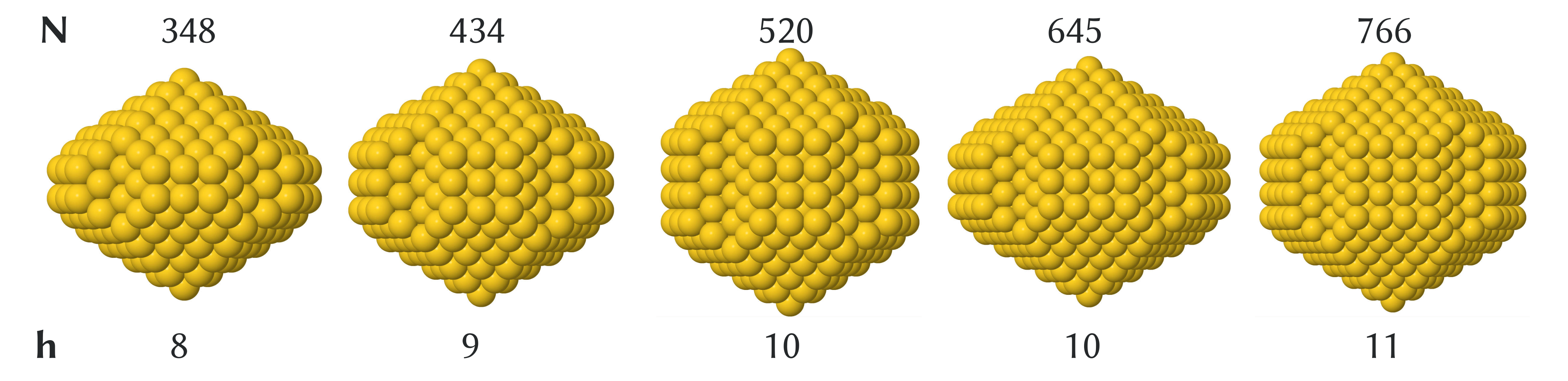}
    \caption{
    Marks decahedra selected for geometry modification. The fivefold axis is parallel to the plane of observation, and the re-entrances that characterize the Marks geometry are clearly visible. The size ranges from $N=348$ to 766 atoms, with axis heights varying between 8 and 11 atoms.}
    \label{fig:5deca}
    \includegraphics[width=.8\textwidth]{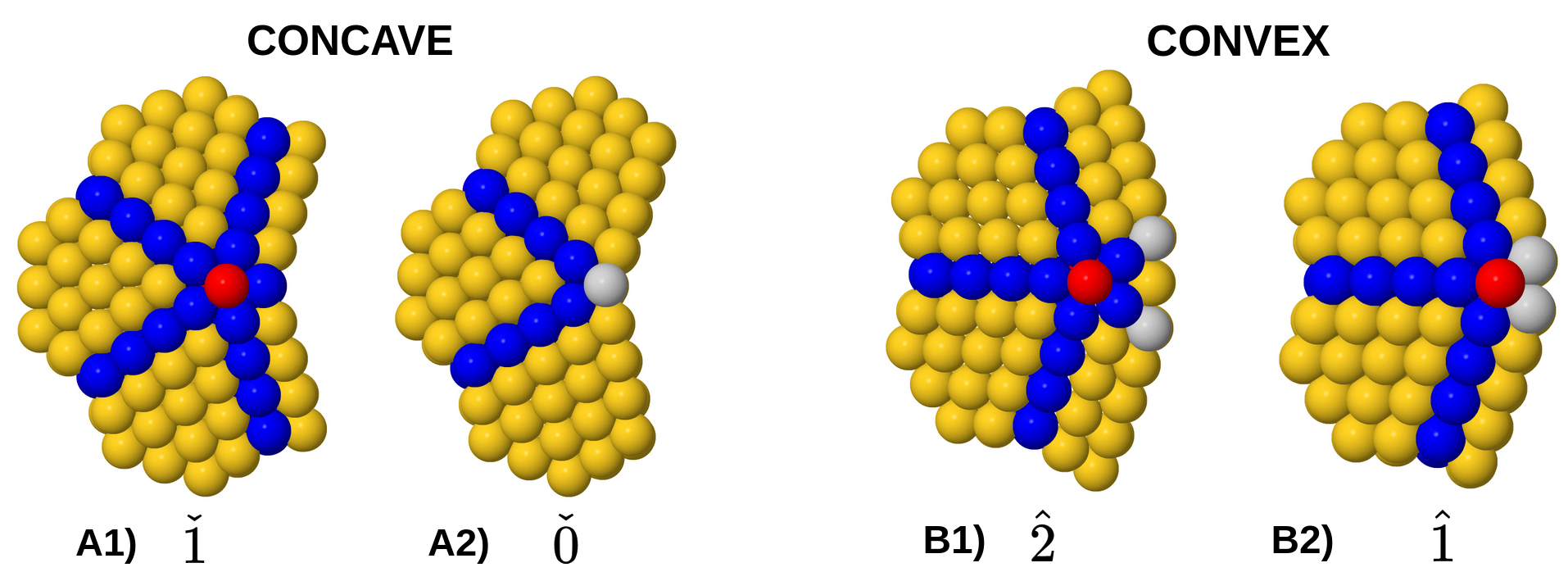}
    \caption{Asymmetric removal of atomic layers from the Dh348 cluster. The structures are oriented with the fivefold axis perpendicular to the plane of observation. A1 and A2 depict concave ($\check{n}$) geometries with the axis beneath one and zero layers, respectively. B1 and B2 depict convex ($\hat{n}$) geometries with the axis beneath two and one layers. Atoms are colored according to their local environment: blue indicates HCP twin planes, red indicates the fivefold axis.
    }
    \label{fig:concave_and_convex_cuts}
\end{figure}

We consider five stable Marks decahedral clusters, ranging from $N=348$ to $766$ atoms and exhibiting different aspect ratios (see \autoref{fig:5deca}), and
we modify their geometry. Specifically, we asymmetrically remove outer atomic layers to generate two distinct classes of structures, concave and convex, examples of which are shown in \autoref{fig:concave_and_convex_cuts}. In the following, we use the notation $\check{n}$ to indicate concave geometries and $\hat{n}$ to denote convex ones, where the axis lies beneath $n$ atomic layers. These modifications reduce symmetry and compactness and bring the fivefold axis closer to the surface, thereby lowering energetic stability and facilitating the transformation of topological defects on nanosecond timescales.

Starting from these modified clusters, we perform molecular dynamics simulations of 1 \textmu s duration at constant temperature, combined with common‑neighbour analysis and local pressure calculations.

We investigate the mechanisms governing the stability of fivefold twins, focusing on the competition between axis centering and detwinning. Our results demonstrate that surface geometry determines these evolutionary pathways: convex structures facilitate the annihilation of the fivefold axis, whereas surface concavity promotes axis centering or even its nucleation through collective atomic displacements.

\section{Model and methods}
\subsection{Atomistic potential}
To model the interactions between gold atoms, this study utilizes an atomistic potential derived from the second-moment approximation of the tight-binding model, developed by Gupta and Rosato \textit{et al.} \cite{Rosato1989,Gupta1981}, commonly referred to as the Gupta potential. The potential represents the total energy $E_i$ of an atom $i$ as the sum of a repulsive pair-wise interaction and a many-body attractive term accounting for the local atomic environment, which implicitly incorporates electronic effects: 
\be  
\label{gupta_potential}
E_i =   \sum_{j, r_{ij} < r_c} A \exp \left[ -p \left( \frac{r_{ij}}{r_0} - 1 \right) \right] - \sqrt{\sum_{j, r_{ij} < r_c} \xi^2 \exp \left[ -2q \left( \frac{r_{ij}}{r_0} - 1 \right) \right]} \ ; 
\ee
in this expression, $r_{ij}$ is the distance between atoms $i$ and $j$, $r_c$ is the cutoff radius of the interactions and $r_0$ is the nearest-neigbor distance in the bulk.
In general, the parameters of the Gupta potential are fitted to reproduce selected experimental properties of bulk metals, and different fitting strategies may lead to distinct parameterizations optimized for different physical features. Different choices of the interaction cutoff further contribute to the diversity of available parameterizations. As a result, several Gupta parameter sets for gold exist in the literature, each reflecting specific fitting choices.

In this work, we employ for the first time a newly developed parameterization, which is explicitly tested and compared against a widely used reference parameter set taken from the literature \cite{Baletto2002}, hereafter referred to as JCP2002. Unlike standard parameterizations that aim to reproduce a limited subset of bulk properties exactly, the present fitting strategy does not enforce an exact match of individual quantities. Instead, the parameters $A, p, q,$ and $\xi$ (reported in \autoref{tab:gupta_parameters}) are determined by simultaneously considering a broader set of experimental bulk properties, namely the equilibrium lattice constant $a_0$, the cohesive energy per atom $E_c$, the bulk modulus $B$, the three independent elastic constants $C_{11}$, $C_{12}$, $C_{44}$, and the energy differences $\Delta E_{\text{FCC-HCP}}$ and $\Delta E_{\text{FCC-BCC}}$, seeking an overall optimal compromise among them. The latter two are particularly relevant for the present study, because they determine the relative stability of local coordinations and the energetic cost of lattice distortions. Regarding the cutoff, our parameterization uses the form of the potential in \autoref{gupta_potential} up to the third-neighbour distance in the bulk crystal, and smoothly links the potential to zero at the fourth-neighbour distance. By contrast, the JCP2002 potential employed a cutoff between the second and third neighbours.

\begin{table}[h!]
\small
    \centering
    \begin{tabular}{l l l}
        \hline
        \textbf{Parameter} & {\textbf{JCP2002} \cite{Baletto2002}}& \textbf{NEW}  \\
        \hline
        $A $ [eV]& 0.2197 & 0.2080 \\
        $\xi $ [eV] & 1.855 & 1.822 \\
        $p$ & 10.53 & 10.0913 \\
        $q$ & 4.3 & 3.9447 \\
        Cutoff start [\AA]& 4.07  &  4.97 \\
        Cutoff end [\AA]&  4.98 & 5.74 \\
        \hline
    \end{tabular}
    \caption{Parameters of the Gupta potential used in this work. Two parameter sets are employed, the NEW parametrization being specifically developed for the scopes of this work. Note that the JCP2002 model truncates interactions between the second and third nearest neighbors, while for the NEW parametrization the cutoff lies between the third and fourth.}
        \label{tab:gupta_parameters}
\end{table}

\begin{table}[ht!]
\small
    \centering
    \begin{tabular}{lllll}
        \hline
        \textbf{Property} & \textbf{Unit} & {\textbf{Experimental}} & {\textbf{JCP2002}} \cite{Baletto2002}& \textbf{NEW}\\
         \hline
        $r_0$ & [\r{A}] & 2.8747 & 2.878 & 2.870 \\
        $a_0$ & [\AA] & 4.0654 & 4.071 & 4.059 \\
        $E_c$ & [eV] & -3.779 & -3.818 & -3.872 \\
        $C_{11}$ & [Mbar] & 1.87 & 2.07 & 1.89 \\
        $C_{12}$ & [Mbar] & 1.55 & 1.72& 1.55 \\
        $C_{44}$ & [Mbar] & 0.45 & 0.45 & 0.46 \\
        $C_p$ & [cal/K mol] & 6.05 & 6.27 & 6.23\\
        $\alpha$ & $[1/\text{K}]$ & 1.42$\cdot10^{-5}$ & 2.39$\cdot10^{-5}$ & 2.17$\cdot 10^{-5}$ \\
        $E_v$ & [eV] & 0.9 & 0.622 & 0.717 \\
        $T_m$& [K] & 1336 & 1110 & 1209 \\
        $\Delta E_{\text{FCC-HCP}}$ & [eV] & 0.005 & -0.0088 & 0.0012 \\
        $\Delta E_{\text{FCC-BCC}}$ & [eV] & 0.04 & 0.027 & 0.037 \\
        $B$ & [Mbar] & 1.66 & 1.82 & 1.66 \\
         \hline
    \end{tabular}
        \caption{Comparison of different physical properties of the Au bulk crystals: experimental vs. calculated with the two parametrizations of the Gupta potential. Overall, the new potential shows improved agreement with experiments with respect to \cite{Baletto2002}. Notably, the FCC–HCP energy difference now has the correct sign, and the melting temperature has increased, moving closer to the experimental value.}
        \label{tab:gupta_full_data_v2}
    \end{table}

\begin{table}[ht!]
\small
\centering

\begin{tabular}{lll}
\textbf{Au38} \\
\hline
Structure & $\Delta$JCP2002 (eV) & $\Delta$NEW (eV) \\
\hline
FCC (TO) & 0 & 0 \\
Dh & 0.0025 & 0.0978 \\
\hline
\end{tabular}
\vspace{.3cm}

\begin{tabular}{lll}
\textbf{Au55} \\
\hline
Structure & $\Delta$JCP2002 (eV) & $\Delta$NEW (eV) \\
\hline
FCC (TO) & 0 & 0 \\
Single-twin & 0.038 & 0.0341 \\
Ih & 0.7210 & 0.2543 \\
Ih rosette & 0.0866 & 0.0755 \\
\hline
\end{tabular}
\vspace{.3cm}

\begin{tabular}{lll}
\textbf{Au147} \\
\hline
Structure & $\Delta$JCP2002 (eV) & $\Delta$NEW (eV) \\
\hline
Dh 146+1 & 0 & 0 \\
Single-twin & 0.047 & 0.2057 \\
FCC & 0.1089 & 0.1580 \\
Ih rosette & 0.9104 & 0.8309 \\
Ih & 1.8649 & 1.1894 \\
\hline
\end{tabular}
\vspace{.3cm}

\begin{tabular}{lll}
\textbf{Au201} \\
\hline
Structure & $\Delta$JCP2002 (eV) & $\Delta$NEW (eV) \\
\hline
FCC (TO) & 0 & 0 \\
Dh & 0.0524 & 0.1324 \\
Single-twin & 0.0677 & 0.0907 \\
\hline
\end{tabular}
\vspace{.3cm}

\begin{tabular}{lll}
\textbf{Au294} \\
\hline
Structure & $\Delta$JCP2002 (eV) & $\Delta$NEW (eV) \\
\hline
Dh & 0 & 0 \\
FCC & 0.2163 & 0.3576 \\
Single-twin & 0.2221 & 0.2992 \\
Ih & 1.8732 & 2.0337 \\
\hline
\end{tabular}
\vspace{.3cm}

\begin{tabular}{lll}
\textbf{Au309} \\
\hline
Structure & $\Delta$JCP2002 (eV) & $\Delta$NEW (eV) \\
\hline
FCC (TO) & 0 & 0 \\
Dh & 0.0139 & 0.0200\\
Ih rosette & 2.0755 & 1.959 \\
Ih & 3.7870 & 2.7498 \\
\hline
\end{tabular}
\vspace{.3cm}

\begin{tabular}{lll}
\textbf{Au561} \\
\hline
Structure & $\Delta$JCP2002 (eV) & $\Delta$NEW (eV) \\
\hline
Dh & 0 & 0 \\
Single-twin & 0.0948 & 0.2299 \\
Ih rosette & 4.4640 & 4.1260 \\
Ih & 6.5097 & 4.9925 \\
\hline
\end{tabular}

\caption{Potential energies of the lowest-energy structures belonging to different geometric motifs, as obtained by BH global optimization searches with the JCP2002 \cite{Baletto2002} and NEW  parameterizations. For each size, energies are calculated with respect to the energy of the global minimum, which is set to zero. The best structures predicted by the two parameterizations match. The energy of icosahedral clusters is high for both parameterizations, but for the new one is slightly lower. Experimentally, a greater number of icosahedral clusters are observed than theory predicts; this trend is partially reproduced by our potential.}
\label{tab:best_struct_comparison}
\end{table}
\normalsize

\begin{table}[h!]
\small
    \centering
    
    \begin{tabular}{@{\extracolsep{\fill}}lllllll}
        \textbf{$\text{Au309}$} \\
        \hline
       Potential & FCC1 & FCC2 & FCC3 & Dh1 & Dh2 & Dh3  \\
        \hline
        \textbf{JCP2002} \cite{Baletto2002} & 600 & 605/610 & 600/605 & 620 & 630 & 630  \\
        \textbf{NEW} & 640 & 660 & 670 & 675 & 655 & 685 \\
        \hline
    \end{tabular}

    \vspace{.2cm} 

    
    \begin{tabular}{@{\extracolsep{\fill}}lllllll}
        \textbf{$\text{Au561}$} \\
        \hline
        Potential & Dh1 & Dh2 & Dh3 & S-twin1 & S-twin2 & S-twin3 \\
        \hline
         \textbf{JCP2002} \cite{Baletto2002} & 690 & 685 & 685 & 665 & 665 & 670/675 \\
        \textbf{NEW} & 740 & 740 & 745 & 730 & 720 & 725 \\
        \hline
    \end{tabular}
  
    \vspace{.2cm} 
    
    \begin{tabular}{@{\extracolsep{\fill}}llll}
    \textbf{$\text{Au586}$} \\
        \hline
Potential & FCC1 & FCC2 & FCC3 \\
        \hline
         \textbf{JCP2002} \cite{Baletto2002}& 680 & 680 & 683 \\
        \textbf{NEW} & 730 & 735 & 740 \\
        \hline
    \end{tabular}
  
        \caption{Melting temperatures $T_{\text{m}}$ (K) of gold clusters as obtained by MD simulations with the two different parameterizations, and for different morphologies. Experimental melting temperatures for clusters of 309 and 561 atoms are approximately 670 K and 770 K, respectively, as reported in \cite{Foster2019}, after accounting for the carbon support that suppresses the melting point.}
    \label{tab:au_cluster_melting}
\end{table}

The reliability of the newly developed parameterization is assessed through a systematic comparison with the reference Gupta potential JCP2002 and with available experimental data, at both the bulk and cluster levels.

At the bulk level, both parameterizations are used to compute structural, elastic, and energetic properties of crystalline gold. The new parameterization shows an overall improved agreement with experimental data. Crucially, it yields the correct physical sign for the FCC–HCP energy difference ($\Delta E_{\text{FCC-HCP}}$), which is a necessary condition for reliably modeling twin-boundary formation and annihilation, since this quantity governs the relative stability of local atomic coordinations and the energetic cost of lattice distortions (see \autoref{tab:gupta_full_data_v2}). We note that the choice of the cutoff is crucial in this sense. If the cutoff is too short, the FCC and HCP phases are not well distinguished (they are identical up to the first-neighbour distance), and the potential may not capture the energetic differences between the two different stacking sequences, which is the case of the JCP2002 parametrization. Extending the cutoff to include interactions up to the fourth-neighbour distance allows the model to correctly reproduce the physical sign of the FCC–HCP energy difference.

At the cluster level, the JCP2002 parameterization is known to provide a very good description of the energetically most stable geometries across a wide range of sizes and morphologies, in agreement with experimental observations for free and supported clusters \cite{Wells2015,Han2014,WangPalmer2012} and first-principles calculations for free clusters \cite{Palomares2017}. However, despite its success in reproducing cluster geometries, the JCP2002 potential exhibits two main limitations that are relevant for the present work. First, it systematically underestimates the melting temperatures of gold clusters of all considered sizes \cite{Foster2019}, a deficiency that can be traced back to the corresponding underestimation of the bulk melting temperature. Second, it tends to over-stabilize FCC-based fragments and decahedral motifs relative to icosahedral structures, making icosahedra excessively unfavorable from an energetic standpoint. This behavior is at odds with experimental observations, which report a significantly larger fraction of icosahedral clusters than predicted by this model \cite{Dearg2024}.

The new parameterization alleviates both of these shortcomings. Basin Hopping global optimization searches show that it preserves the ability of the JCP2002 potential to correctly identify the lowest-energy structural motifs across the investigated cluster sizes (see \autoref{tab:best_struct_comparison} for a direct comparison of the best-predicted structures). At the same time, molecular dynamics simulations of cluster melting reveal a systematic increase in the predicted melting temperatures, leading to values in significantly better agreement with experimental measurements (see \autoref{tab:au_cluster_melting}). This improvement is observed consistently for all cluster sizes considered ($N=309, 561, 586$) and provides a more robust basis for simulations of solid-state structural evolution. In addition, the energetic separation between icosahedral clusters and competing FCC or decahedral structures remains sizable but is reduced with respect to the JCP2002 parameterization, resulting in a more realistic relative stability of different morphologies (see again \autoref{tab:best_struct_comparison}).

\subsection{Methods}
The evolution of the clusters is studied by molecular dynamics (MD) simulations using our own MD code, in which equations of motion are solved by the velocity Verlet algorithm with a time step of 5 fs. We perform simulations at constant temperature, \textit{i.e.} in the $NVT$ ensemble, of  1\ \textmu s duration, in order to observe slow atomic rearrangements. The temperature is kept constant by using the
Andersen thermostat \cite{li2005}. The setting of the thermostat is done as in \textit{ref.} \cite{Baletto2000_surf}.  We also perform MD heating simulations for each cluster size to determine their melting temperatures, so we can choose temperatures for evolution simulations safely below the melting points. Heating is carried out increasing temperature at a rate of 1 K/ns. After each run, we perform L-BFGS optimization to locate the local minimum corresponding to each saved frame for subsequent structural and energetic analysis.

Global optimization searches are used to locate the lowest energy structures for the cluster sizes considered in this study. These searches are performed by the Basin Hopping (BH) algorithm \cite{Wales1997} as implemented in our code \cite{Rossi2009}. 

Structural motifs are singled out by using the common neighbour analysis (CNA) signatures \cite{Faken1994} of nearest-neighbour (nn) atomic pairs. 
In the CNA, to each nn atom pair a three‑integer signature $(m, n, p)$ is assigned, where $m$ denotes the number of common nn of the two atoms, $n$ the number of nn bonds among these common neighbours, and $p$ the maximum length of a chain that can be formed with the bonds.
Specifically, the fraction of nn pairs with (555) signature allows to distinguish between icosahedral, decahedral and crystalline structures, while the fraction of (422) signatures allows to discriminate between purely FCC clusters from those presenting defects such as stacking faults and twin planes. A table with the typical fractions of (555) and (422) signatures in this size range can be found in ref. \cite{Faken1994}. Furthermore, CNA allows for the precise classification of atoms into distinct categories according to the signatures calculated with all the pairs formed with its nn  \cite{Roncaglia2023}. Atomic categories relevant for this study are reported in \autoref{tab:cna}.

\begin{table}[ht!]
\small
    \centering
    \begin{tabular}{lll}
        \hline
        nn & list of signatures & name\\
        \hline
        12 & 421 421 421 421 421 421 421 421 421 421 421 421 & FCC \\
        12 & 421 421 421 421 421 421 422 422 422 422 422 422 & HCP \\
        12 & 422 422 422 422 422 422 422 422 422 422 555 555 & 5-axis \\
        \hline
    \end{tabular}
    \caption{Each atom can be classified according to the CNA signatures computed with all its neighbours. In this work we mainly use three classes derived from these signatures to distinguish atoms in the FCC (inner FCC), atoms in the twin planes (inner HCP), and atoms belonging to the five-fold axis (inner 5-axis). \cite{Roncaglia2023}. }
        \label{tab:cna}
\end{table}

Another quantity we monitor is the atomic-level pressure, which is derived from the atomic stress tensor \cite{Vitek1987,Nelli2023,Nelli2021}. For every atom within the nanoparticle, the stress tensor is defined as follows 
\be 
\sigma_i^{ab} = \frac{1}{V_i} \sum_{j \neq i} \frac{\partial E_i}{\partial r_{ij}} \frac{r_{ij}^a r_{ij}^b}{r_{ij}}, \label{eq:atom_stress} 
\ee
where $E_i$ represents the atomic energy of the $i$-th atom and $V_i$ denotes its reference volume in the bulk crystal structure. The variables $r^a_{ij}$ and $r^b_{ij}$ (with Cartesian indices $a, b = x, y, z$) correspond to the components of the vector $\mathbf{r}_{ij}$ connecting atoms $i$ and $j$, while $r_{ij}$ represents its magnitude. The isotropic local pressure acting on atom $i$ is proportional to the trace of the tensor:
\be
P_i = -\frac{1}{3} \text{Tr}(\underline{\sigma_i}).
\label{eq:atom_pressure}
\ee
Under this convention, positive values of $P_i$ indicate compressive stress acting on the atom, while negative values indicate tensile stress.

\section{Results}

\subsection{Evolution of concave structures}
Concave geometries are frequently observed during the oriented attachment of gold nanoparticles, where the formation of concave junctions often precedes the emergence of fivefold-twinned configurations \cite{song_oriented_2020}.
We investigate this behaviour for both the $\check{1}$ and $\check{0}$ structures in order to characterise the underlying mechanisms.

\subsubsection*{$\check{1}$ structures (axis beneath 1 layer)}
The clusters obtained by cutting the initial Marks decahedra into $\check{1}$ configurations (see \autoref{fig:concave_and_convex_cuts}) contain 277, 342, 407, 495, and 584 atoms, respectively. Before performing MD evolution simulations, BH global optimization searches are carried out to determine the energetic global minima for these specific sizes. The geometric motifs of the resulting global minima are summarized in \autoref{tab:best_struct_conc1}.
Remarkably, the Dh motif is not the global minimum at all sizes. In two cases (342 and 584 atoms), lower-energy FCC structures and single-twin structures (\textit{i. e.}, structures with a single twin plane separating two FCC regions) are found. This indicates that, if the structure remains decahedral during constant-temperature evolution, it is likely due to kinetic trapping.

\begin{table}[h!]
\small
\centering
\begin{tabular}{llllll}
\hline
& \textbf{Au277} &  \textbf{Au342} & \textbf{Au407} &\textbf{Au495} & \textbf{Au584} \\
\hline
Structure & $\Delta$E(eV)  & $\Delta$E(eV)  & $\Delta$E(eV)  & $\Delta$E(eV)  & $\Delta$E(eV)  \\
\hline
Dh & \textbf{0} & 0.01 & \textbf{0} & 0.15 & 1.20 \\
FCC & 0.13 & 0.45 & 0.64 & \textbf{0} & \textbf{0} \\
Single-twin & 0.14 & \textbf{0} & 0.05 & 0.76 & 0.53 \\
\hline
    
\end{tabular}
\caption{Potential energies of the lowest-energy structures belonging to three different geometric motifs (Dh, Single-twin and FCC), as obtained by BH global optimization searches. Sizes correspond to the five sizes of the $\check{1}$ structures. For
each size, energies are calculated with respect to the energy of the global minimum, which is set to
zero.}
\label{tab:best_struct_conc1}
\end{table}

To select appropriate simulation temperatures, we estimate the melting points ($T_m$) via heating simulations. We then perform MD simulations at constant temperatures strictly below $T_m$ to observe solid-state evolution. For each structure, 10 independent evolution simulations are carried out (five per temperature, at at least two different temperatures).

The evolution results demonstrate the persistence of the fivefold-twin motif. Regardless of the global minimum for the corresponding size, the concave configurations consistently evolve towards Dh clusters, as shown in \autoref{tab:evolution_conv1}. Only in two simulations at 625 K the fivefold axis is lost, leading to the formation of a structure with only one twin plane (single-twin).

\begin{figure}[h!]
    \centering
     \includegraphics[width=0.9\textwidth]{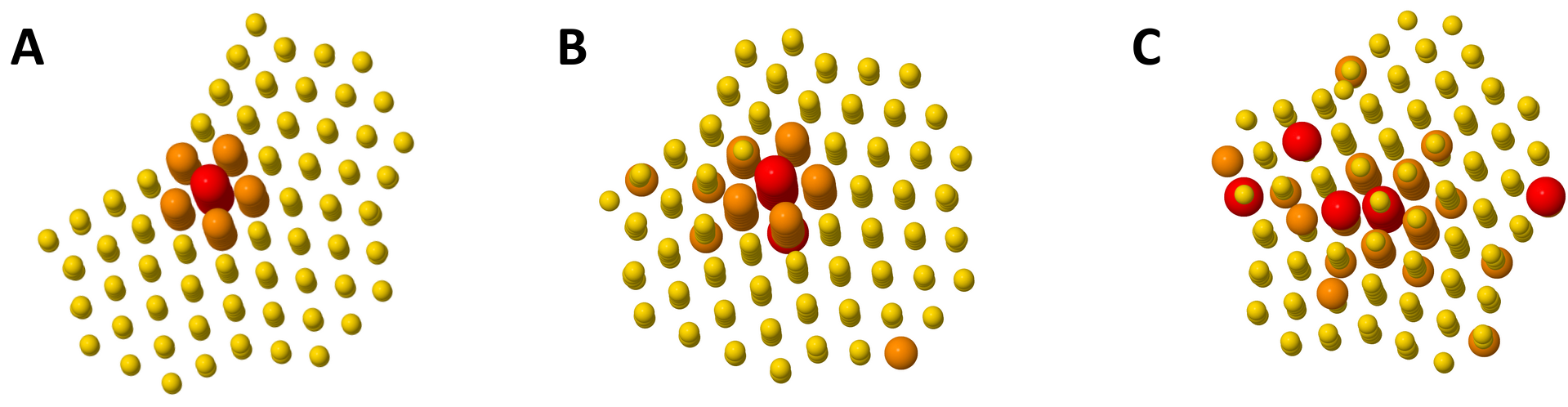}
     \caption{Time-sequence of the recentering process in a 277-atom $\check{1}$ cluster at 600 K. The disclination axis (red) and its first-neighbor columns (orange) are stationary while surface atoms migrate toward the re-entrant facets. A) Initial peripheral axis (0 ns); B) intermediate filling of the concavity (8 ns); C) final restored fivefold symmetry (1\ \textmu s).}
    \label{fig:348_conv1_md4_600K}
\end{figure}

The evolution within the decahedral motif is primarily driven by collective surface diffusion. The atomic columns constituting the fivefold axis and its immediate surroundings exhibit strong structural persistence, largely retaining their lattice positions (see red and orange atoms in \autoref{fig:348_conv1_md4_600K}). In contrast, surface atoms undergo directed mass transport that progressively fills the concavity, shifting the nanoparticle's center of mass and resulting in the effective recentering of the fivefold axis.

\begin{figure}[h!]
    \centering
    \includegraphics[width=0.9\textwidth]{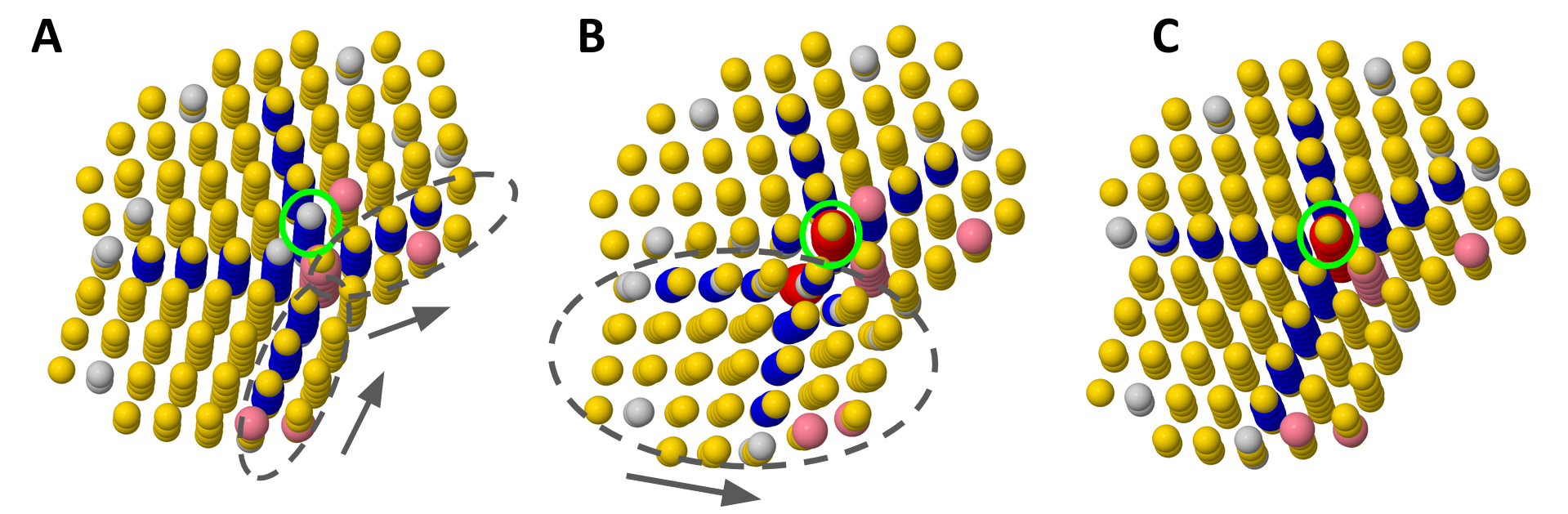}
    \caption{fivefold axis migration during the evolution at 550K of the $\check{1}$ structure obtained from Dh434 captured between 388.78 and 388.80 ns. A) The axis is mainly composed of the same atoms as the original one (colored in pink); in B) we observe a collective movement of groups of columns which leads to the nucleation of a new fivefold axis (red atoms) in C), distinct from the original one and located in a more central layer.}
    \label{fig:evoconcave434_1Lzoom}
\end{figure}

Less frequently, an alternative mechanism drives the recentering process through the nucleation of a new fivefold axis, which replaces the original one. In this regime, collective atomic displacements induce the formation of a parallel axis along a more central column, within one of the original twin planes (see \autoref{fig:evoconcave434_1Lzoom}). While the preservation of the original axis is the exclusive outcome in 277- and 495-atom clusters, the nucleation of a parallel axis occurs in a minority of cases for the 342- and 584-atom clusters, and is observed in 4 out of 14 simulations for the 407-atom structure.

Both processes are thermodynamically driven by the reduction of the unfavorable surface-to-volume ratio associated with the initial concave geometry, leading to a decrease in total surface energy and a partial restoration of global fivefold symmetry.

\begin{table*}[h!]
\small
\centering
\begin{tabular}{clccccccc}
%
\hline
 size &best& $T_m$ &T = 550 K & T = 575 K & T = 600 K & T = 625 K & T = 650 K  & total \% \\
\hline 
\textbf{277} &Dh & $\sim$ 660&  100\% & -- & 100\% & -- &-- &100\% \\
\textbf{342} &FCC & $\sim$ 675& 100\%  & -- & -- & 80\% & -- &90\%\\
\textbf{407} &Dh &$\sim$ 710& -- & 100\% & 100\% & 80\% & -- & 93\%\\ 
\textbf{495} &Dh &$\sim$ 725& -- & 100\% & -- & -- &100\% & 100\%\\ 
\textbf{584} &S-twin& $\sim$ 730& -- &  -- &  100\% & -- & 100\%  & 100\% \\

\hline
    
\end{tabular}
\caption{Percentage of decahedra at the end of constant-temperature MD simulations (after 1 $\mu $ s), depending on size and temperature}
\label{tab:evolution_conv1}
\end{table*}

\subsubsection*{$\check{0}$ structures  (axis beneath 0 layer)}
To test the limits of axis stability in concave configurations, we analyze $\check{0}$ structures, in which the cut is deep enough that the axis (no longer properly fivefold) is exposed at the surface. We consider clusters of 234 and 346 atoms, derived from Dh348 and Dh520 respectively (see \autoref{fig:concave_and_convex_cuts}a). These sizes are chosen to probe the effect of axis length (8 versus 10 atoms) on stability; indeed the two clusters are identical in cross-section, and differ only in the length of the atomic columns.
As for the $\check{1}$ structures, BH global optimization searches are performed to identify the global minima for these sizes (see \autoref{tab:best_struct_conc0}), and heating simulations are used to estimate the corresponding melting temperatures.
\begin{table}[h!]
\small
\centering
\begin{tabular}{lll}
 & \textbf{Au234} &  \textbf{Au346} \\
\hline
Structure & $\Delta$E(eV)  & $\Delta$E(eV) \\
\hline
Dh & 0.03 & \textbf{0} \\
FCC & 0.03 & 0.09 \\
single-twin & \textbf{0} & 0.35 \\
\hline   
\end{tabular}
\caption{Potential energies of the lowest-energy structures belonging to three different geometric motifs
(Dh, Single-twin and FCC), as obtained by BH global optimization searches. Sizes correspond to the sizes of the $\check{0}$ structures. For each size, energies are calculated with respect to the energy of the
global minimum, which is set to zero.}
\label{tab:best_struct_conc0}
\end{table}

\begin{table}[h!]
\small
\centering
\begin{tabular}{clccccc}
\hline
 size &best &$T_m$& T = 500 K & T = 550 K & T = 600 K \\
\hline 
\textbf{234} &single-twin &$\sim$ 600& 90\% & 80\% & --\\
\textbf{346} &Dh &$\sim$ 675& -- & 60\% &30\% \\
\hline
    
\end{tabular}
\caption{Percentage of decahedra after 500 ns of MD evolution at constant temperature, calculated over ten simulations per system, size and temperature.}
\label{tab:evolution_conv0}
\end{table}

We run 10 MD simulation in the canonical ensemble for each cluster, selecting two different temperatures below the melting ones.
The final structures display a greater diversity than in the $\check{1}$ case, including decahedra, single-twin and fully FCC configurations. Nevertheless, the results reveal a clear driving force toward retwinning and axis centering, even when the original axis is exposed at the surface (see \autoref{tab:evolution_conv0}).
The mechanisms responsible for retwinning and centering of the fivefold axis are similar to those undergone by the $\check{1}$ structure. In most cases, the original axis and twin planes maintain their relative position, and the concavity is progressively filled by diffusing atoms. This mechanism dominates for the 346-atom cluster.

\begin{figure}[h!]
    \centering
    \includegraphics[width=1\linewidth]{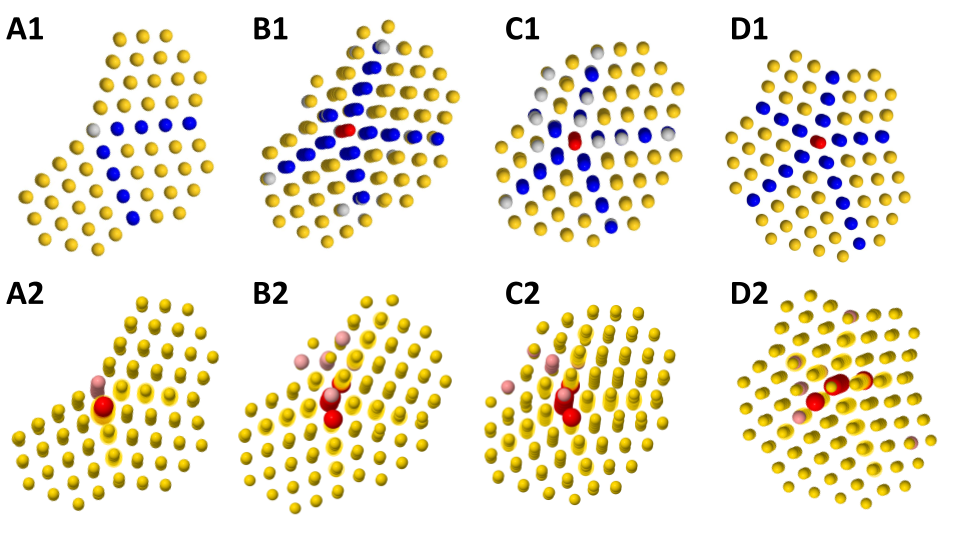}
    
    \caption{
  Evolution of a 234-atom $\check{0}$ cluster at 500 K. In the first sequence, atom belonging to the twin-planes and to the fivefold axis are colored in blue and red, respectively, to highlight the developing decahedral arrangement. In the second sequence, atoms belonging to the original axis are colored in pink, and atoms belonging to the parallel subsurface column that will become the fivefold axis are colored in red. A) Initial unstable exposed axis. B) Subsurface nucleation (9 ns): a new fivefold axis forms in a subsurface column through collective shearing. C) Covering of the new axis (13 ns). D) Final centered decahedron (1\ \textmu s).} 
    \label{fig:evo_concave_0Lnewaxis}
\end{figure}

\begin{figure}[h!]
    \centering
     \includegraphics[width=0.9\textwidth]{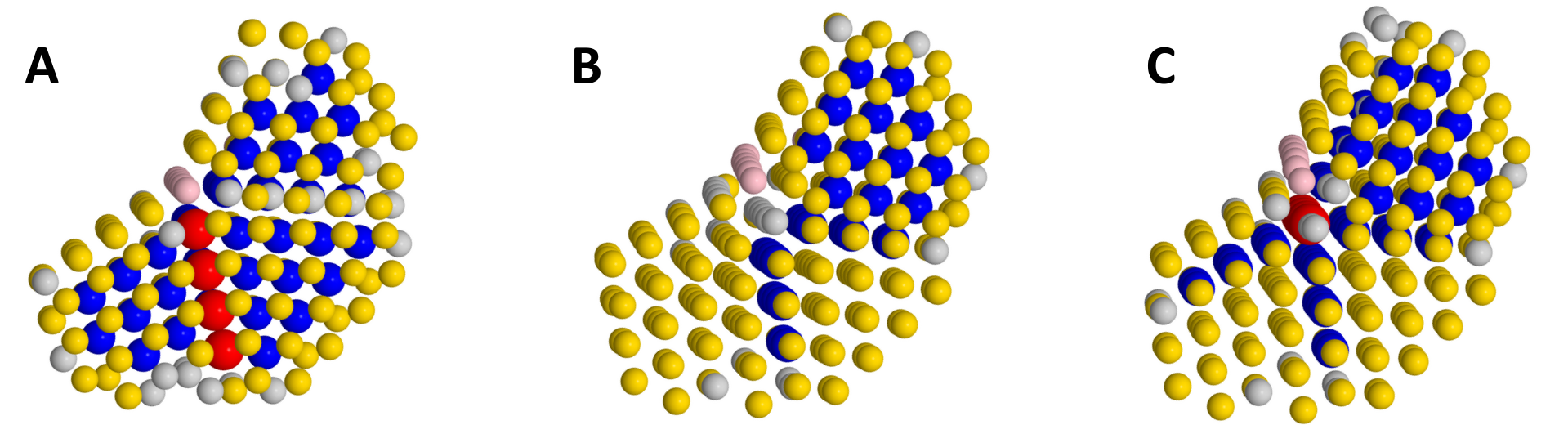}
    \caption{
    Picosecond-scale zoom of the disclination migration in 234-atom $\check{0}$ structure. A) (5.00 ns) Pre-transition state with unstable HCP configurations; B) (5.38 ns) lattice distortion and curving of surface columns; C) (5.40 ns) collective glide of the surface layer leading to the emergence of the new disclination core.}
    \label{fig:348_conv0_md2_zoom}
\end{figure}

For the smaller 234-atom cluster, recentering mechanism such as that of \autoref{fig:evoconcave434_1Lzoom} emerge, characterized by the nucleation of a new fivefold axis. 
As in the $\check{1}$ case, the new axis forms parallel to the original one, and corresponds to an atomic column belonging to one of the original twin planes, typically located one layer beneath the concavity. Once nucleated, the surrounding atoms reorganize to center the new axis.
In \autoref{fig:evo_concave_0Lnewaxis} we show a representative example of the full evolution of the cluster, up to 1 \textmu s, while in \autoref{fig:348_conv0_md2_zoom} the nucleation of the new subsurface fivefold axis is shown in more details. The process occurs on the picosecond timescale through the collective displacement of an entire atomic layer the concave surface. This process can be interpreted as a form of disclination migration, where the topological defect relocates to a more energetically favorable, highly coordinated position (see  \autoref{fig:evo_concave_0Lnewaxis}). 
We note that the outcome of the competition between recentering of the axis and complete detwinning is sensitive to the simulation temperature, with detwinning becoming increasingly likely as temperature rises. For the 346-atom cluster, for instance, at 600 K only 30\% of the simulations lead to restoration of the full fivefold symmetry.
This temperature dependence suggests that while concavity geometrically favors recentering at moderate temperatures, thermal activation progressively lowers the kinetic barriers associated with detwinning, shifting the balance toward the elimination of the fivefold axis at higher temperatures.

\subsection{Evolution of convex structures}

Unlike concave structures, convex geometries ($\hat{n}$) do not exhibit  the re-entrant surface features that favor fivefold symmetry restoration. We investigate how convex configuration influences the detwinning mechanism by studying the evolution of both $\hat{1}$ and $\hat{2}$ structures. 
\subsubsection*{$\hat{1}$ structures (axis beneath 1 layer)}
The clusters obtained by cutting the initial Marks decahedra of \autoref{fig:5deca} into $\hat{1}$ configurations consist of 216, 265, 314, 377, and 443 atoms, respectively. First, we perform BH global optimization searches to determine the global minima (see \autoref{tab:best_struct_convex1}) and we run heating simulation to estimate the melting temperatures of the clusters. 
Then we perform 10 MD simulations in the canonical ensemble at three different temperature below the melting point for each size, of 1 \textmu s duration.
\begin{table}[h!]
\small
\centering
\begin{tabular}{llllll}
\hline
 & \textbf{Au216} &  \textbf{Au265} & \textbf{Au314} &\textbf{Au377} & \textbf{Au443} \\
 Structure& $\Delta$E(eV)  & $\Delta$E(eV)  & $\Delta$E(eV)  & $\Delta$E(eV)  & $\Delta$E(eV)  \\
\hline
Dh & 0.002 & \textbf{0} & 0.23 & 0.11 & \textbf{0} \\
FCC & 0.03 & 0.25 & \textbf{0} & \textbf{0} & 0.22 \\
Twin & \textbf{0} & 0.01 & 0.50 & 0.09 & 0.33 \\
\hline
\end{tabular}
\caption{Potential energies of the lowest-energy structures belonging to three different geometric motifs
(Dh, Single-twin and FCC), as obtained by BH global optimization searches. Sizes correspond to the
sizes of the $\hat{1}$ structures. For each size, energies are calculated with respect to the energy of the global
minimum, which is set to zero.}
\label{tab:best_struct_convex1}
\end{table}

\begin{figure}[h!]
    \centering
    \includegraphics[width=0.95\textwidth]{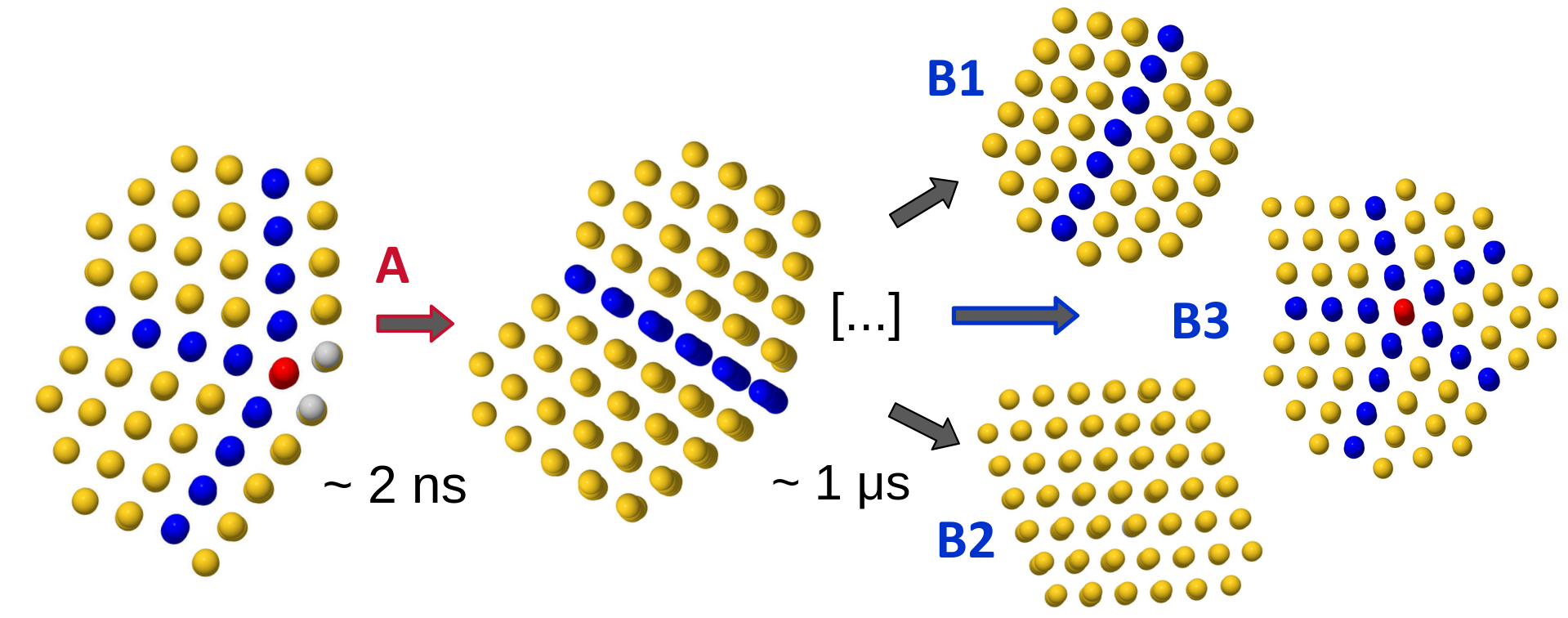}
    \caption{In $\hat{1}$ structures, the detwinning (A) occurs rapidly. Subsequent evolution may preserve the twin plane (B1), proceed to full detwinning to pure FCC (B2),  or re-twin forming a new fivefold axis (B3)}
    \label{fig:evo_conv3481_L}
\end{figure}

The evolution pathways of $\hat{1}$ structures and the resulting final configurations exhibit much greater variety compared to the corresponding concave cases ($\check{1}$ structures).
In all cases, the structure undergoes a rapid detwinning process, occurring in the very first nanoseconds of the simulations. During such process, two of the three original twin planes (and therefore the fivefold axis) disappear. The outcome is a single-twin structure, in which the twin plane splits the cluster in two FCC regions almost equal in size.
After the detwinning, the system follows one of three possible pathways, schematically represented in  \autoref{fig:evo_conv3481_L} and summarized here below:
\begin{enumerate}
    \item \textbf{Twin preservation:} the structure stabilizes in a single-twin configuration.
    \item \textbf{Full detwinning:} the twin plane migrates to the surface and is annihilated, resulting in a pure FCC single crystal.
    \item \textbf{retwinning:} in some cases, stochastic surface rearrangements generate a local concavity, which subsequently nucleates a new fivefold axis, effectively re-establishing the decahedral motif. 
\end{enumerate}

\begin{table*}[h!]
\small
\centering

\begin{tabular}{clccccccc}

\hline
size & best & $T_{melt}$ (K) & 500 K & 525 K & 550 K & 575 K & 600 K& 625 K \\

\hline 

\textbf{216} &S-twin & 590 - 630 & 40\% & 30\% & 80\% & --   & --    & --   \\

\textbf{265} &Dh & 620       & --   & 20\% & 20\% & 50\% & 100\% & --   \\

\textbf{314}&FCC& 650 - 680 & --   & --   & 10\% & 20\% & 0\%   & --   \\

\textbf{377}&FCC& 680       & --   & --   & 10\% & 30\% & 0\%   & 0\%  \\

\textbf{443}&Dh & 700       & --   & --   & 0\%  & 0\%  & 20\%  & 20\% \\

\hline

\end{tabular}
\caption{Final percentage of Dh after 1\ \textmu s evolution as a function of temperature and cluster size, for $\hat{1}$ structures (10 simulation per size per temperature). The results indicate that for sizes where the decahedron is less stable than the FCC structure (e.g., $N=314, 377$), the pathway toward de-twinning is highly favored}

\label{tab:evolution_convex1}

\end{table*}
The frequency of the three evolution pathways varies with temperature and size of the initial $\hat{1}$ structure. The percentage of Dh at the end of each simulation set is reported in \autoref{tab:evolution_convex1}. Though more data would be needed to identify general trends, our results suggest that for sizes where the presence of twin planes is unfavorable (i.e., where FCC structures are lower in energy than Twin and Dh ones), the retwinning evolution pathway is unlikely. In these cases, the twin preservation is frequently observed, and the full detwinning is observed in some cases, especially at the highest simulation temperatures.

\begin{figure}[h!]   
        \centering
        \includegraphics[width=.8\linewidth]{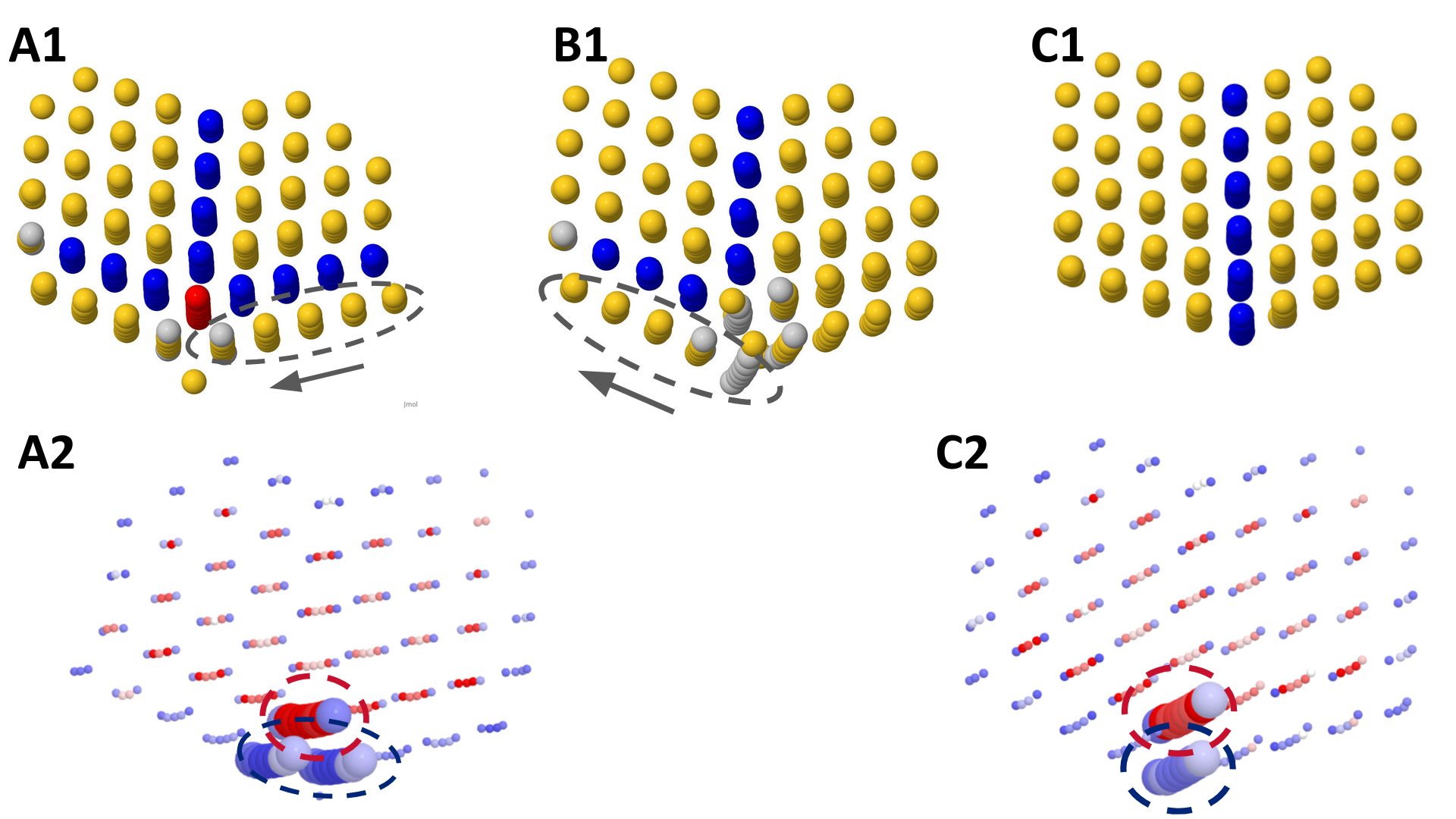}
         \caption{Initial evolution of a $\hat{1}$  structure obtained from Dh348. A) Starting configuration with a highly strained fivefold axis and surface atomic columns. B) Gliding of the surface layer at 1310 ps. C) De-twinned state at 1340 ps with the fivefold symmetry eliminated. Panels A2 and B2 show the atomic-level pressure analysis before and after detwinning, highlighting the reduction in compressive stress and column curvature.}
         \includegraphics[width=.9\linewidth]{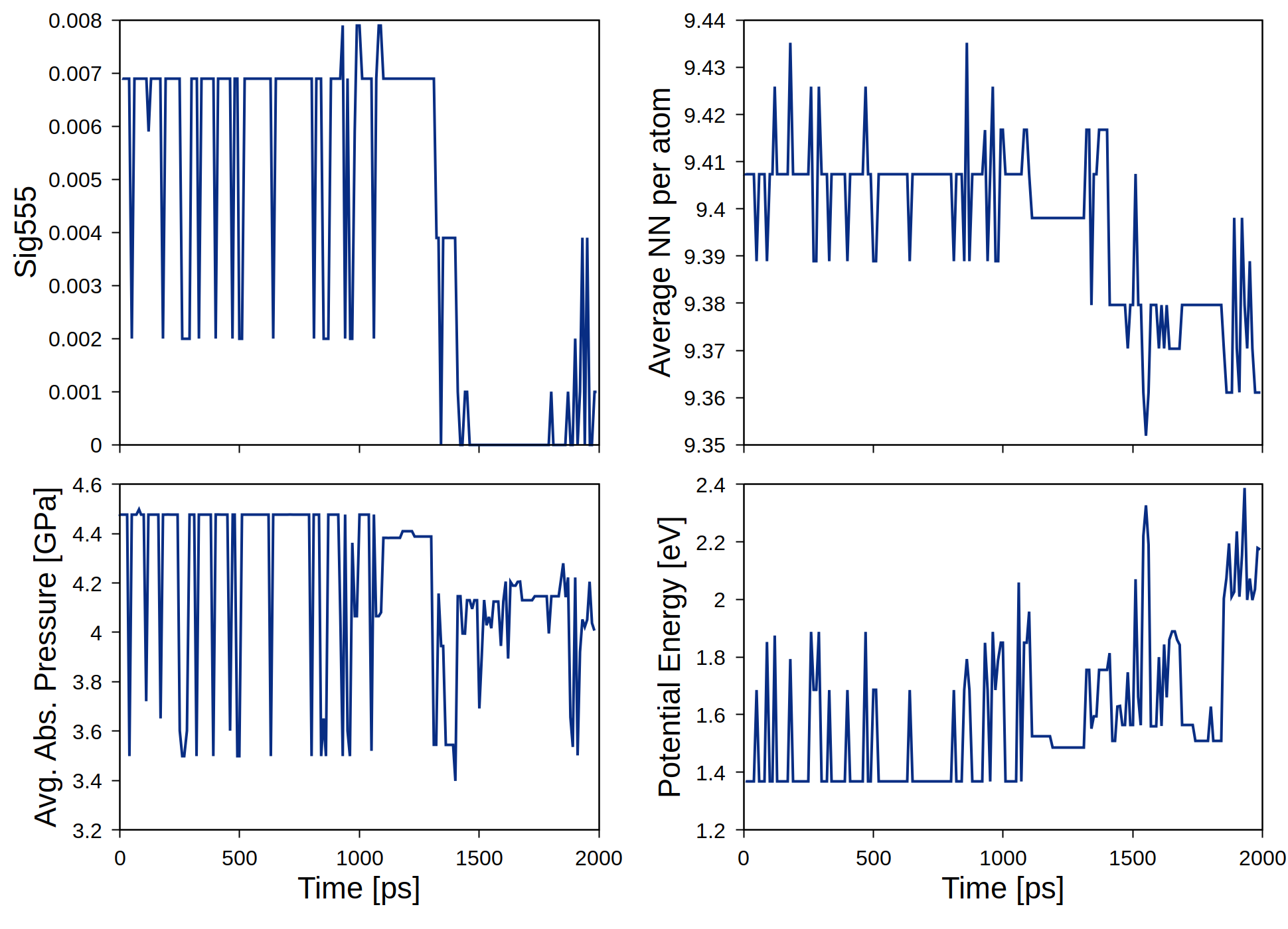}
         \caption{Temporal evolution of key structural and energetic parameters from 0 to 2000 ps. The plots show the sig555 signature tracking the elimination of the fivefold axis, the average number of nearest neighbors per atom, the mean absolute pressure, and the potential energy of the system (with respect to the energy of the global minimum).}        
        \label{fig:evo_348conv1L_detwinning_press}
\end{figure}

In \autoref{fig:evo_348conv1L_detwinning_press}, we show, as an example, the initial detwinning process step-by-step for a $\hat{1}$ structure obtained from Dh348. The analysis of atomic-level pressure reveals that the fivefold axis is under significant compressive stress. Initially, the atomic columns are highly strained and curved, with a mean absolute pressure of approximately 4.39 GPa. The atoms constituting the fivefold axis are misaligned by an average of $0.082$ \AA, while the adjacent front columns exhibit a misalignment of $0.157$ \AA. Stress is relieved through subsequent gliding motions of patches on the convex surface, which begin at 1310 ps. Atoms collectively shift, resulting in a rapid detwinning: by 1340 ps, two twin planes vanish, thereby eliminating the fivefold symmetry. This structural transition, tracked via the sig555 parameter, is accompanied by a reduction in mean absolute pressure to 4.16 GPa (see \autoref{fig:evo_348conv1L_detwinning_press}), alongside decreased column curvature and reduced atomic misalignments ($0.075$ \AA \ for the former fivefold axis and $0.117$ \AA \ for the adjacent columns).

The temporal evolution of the potential energy of the cluster, which is plotted in \autoref{fig:evo_348conv1L_detwinning_press} as well, shows that the initial decahedral motif with an off-centered axis is energetically more favorable than the final single-twin structure. This can be attributed to the reduced compactness of the structure, as evidenced by a decrease in the average number of nearest neighbors per atom (see again \autoref{fig:evo_348conv1L_detwinning_press}).
As a consequence, the system exhibits structural fluctuations between the fivefold and detwinned states over the first 2000 ps before ultimately relaxing into a stable single-twinned particle.
Analogous detwinning mechanisms and structural fluctuations are also observed for the other $\hat{1}$ structures. Average detwinning times are reported in \autoref{tab:detwinning_times_expanded}.

\begin{table}[h!]
\centering
\begin{tabular}{c|cc|cc|cc|cc|cc|cc}
\toprule
& \multicolumn{2}{c|}{$T$ = 500 K} & \multicolumn{2}{c|}{$T$ = 525 K} & \multicolumn{2}{c|}{$T$ = 550 K} & \multicolumn{2}{c|}{$T$ = 575 K} & \multicolumn{2}{c|}{$T$ = 600 K} & \multicolumn{2}{c}{$T$ = 625 K} \\
Size & $t_1$ & $t_d$ & $t_1$ & $t_d$ & $t_1$ & $t_d$ & $t_1$ & $t_d$ & $t_1$ & $t_d$ & $t_1$ & $t_d$ \\
\midrule
\textbf{216} & 6.80 & 30.19 & 4.70 & 74.61 & 1.83 & 19.90 & -- & -- & -- & -- & -- & -- \\
\textbf{265} & -- & -- & 2.873 & 25.510 & 3.400 & 11.700 & 0.890 & 23.812 & 0.52 & 51.45 & -- & -- \\
\textbf{314} & -- & -- & -- & -- & 1.66 & 11.62 & 1.14 & 23.94 & 1.44 & 15.19 & -- & -- \\
\textbf{377} & -- & -- & -- & -- & 3.60 & 7.50 & 1.41 & 15.99 & 1.60 & 20.36 & 1.40 & 5.60 \\
\textbf{443} & -- & -- & -- & -- & 3.40 & 3.80 & 1.80 & 13.20 & 1.40 & 16.20 & 0.70 & 32.00 \\
\bottomrule
\end{tabular}
\caption{Average initial ($t_1$) and definitive ($t_d$) detwinning times (in ns) as a function of nanoparticle size and temperature.}
\label{tab:detwinning_times_expanded}
\end{table}

\begin{figure}[h!]
    \centering
    \includegraphics[width=1\linewidth]{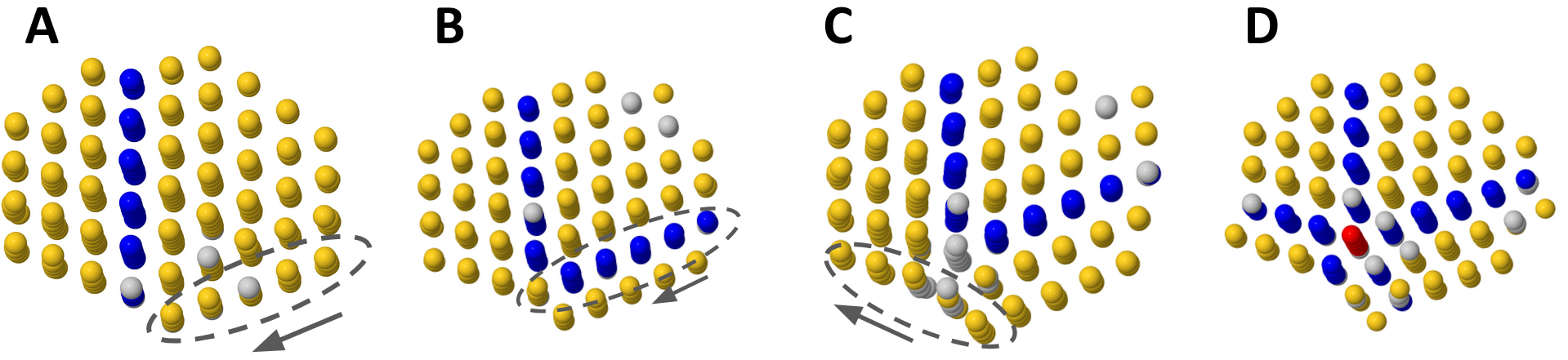}
 \caption{Snapshots illustrating the retwinning mechanism in a 216-atom $\hat{1}$ convex cluster. A) Starting state with a single residual twin boundary (blue). B--C) Collective movement of atomic columns (indicated by dashed lines and arrows) driven by surface fluctuations. D) Nucleation of a new fivefold disclination axis (red) at the junction of newly formed twin planes, restoring the decahedral motif. We can observe a local concavity at the nucleation site, which acts as a preferential site by lowering the energetic barrier for the formation of the new fivefold junction.}
    \label{fig:evo_348conv1L_twinning}
\end{figure}

After the initial detwinning process, one possible pathway is the retwinning, which involves a spontaneous restoration of the decahedral symmetry (see \autoref{tab:evolution_convex1} for statistics). Starting from the single-twinned configuration, thermal fluctuations drive surface diffusion, which in specific instances generates a local surface concavity that functions as a nucleation site. Through the collective displacement of atomic columns, a new fivefold axis forms and is progressively centered via mechanisms analogous to those observed in concave structures. This process, which recovers the global decahedral motif, is illustrated in \autoref{fig:evo_348conv1L_twinning}.

\subsubsection*{$\hat{2}$ structures (axis beneath 2 layers)}
\begin{figure}[h!]
    \centering
    \includegraphics[width=1\linewidth]{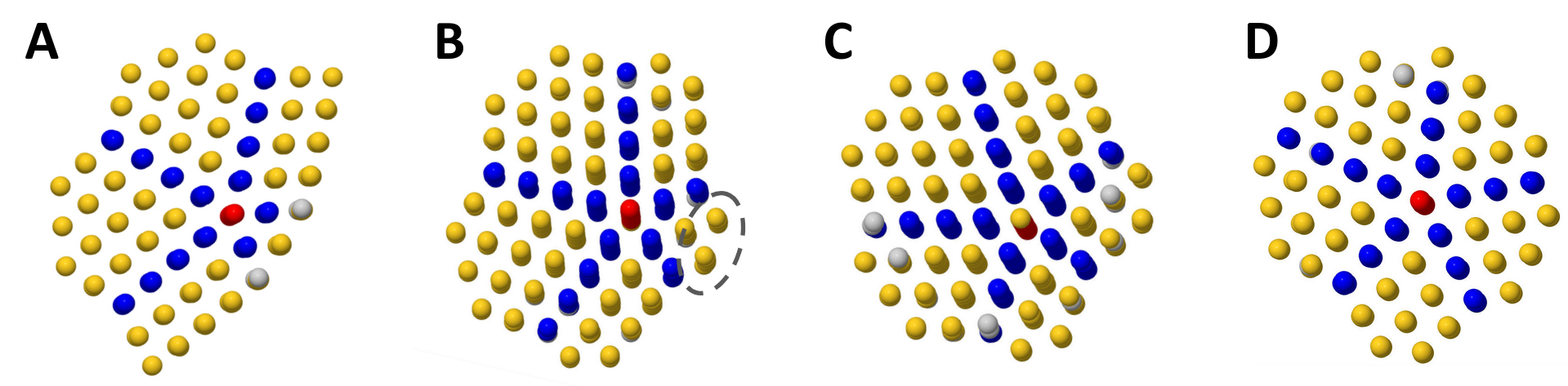}
    \caption{Evolution at $T$ = 550 K of the 262-atom $\hat{2}$ cluster. A) Starting configuration; B) at 10 ns, surface atom diffusion initiates the formation of a secondary layer, starting from the two highlighted columns; C) (500 ns) the structure is more symmetric and compact; D) (1 \ \textmu s) the original axis is centered.}
    \label{fig:evoconv348_2L}
\end{figure}

To investigate whether the detwinning mechanism also occurs when the axis is farther from the surface, we consider the $\hat{2}$ structures. 
As for the $\check{0}$ case, we choose to focus on $\hat{2}$ structures obtained by cutting Dh348 and Dh520, respectively formed by 262 and 382 atoms. After estimating their melting temperature and the most favorable configurations for these sizes , we perform 10 $NVT$ MD simulations at two different temperatures for each structure. Remarkably, detwinning never occurs in any of the simulations performed. Instead, these clusters consistently undergo recentering of the fivefold axis (see \autoref{fig:evoconv348_2L} and \autoref{tab:evolution_convex2} for statistics). This demonstrates that for the sizes investigated the structural outcome changes dramatically when the fivefold axis is positioned beneath two surface layers ($\hat{2}$), rather than one ($\hat{1}$). In these clusters, the additional atomic layer provides sufficient mechanical confinement to suppress the surface glide observed in $\hat{1}$ structures. 
In the Au262 cluster, the original axis is preserved and re-centered (see \autoref{fig:evoconv348_2L}). 

Also the evolution of the Au382 structure always converge to a decahedron, but, similarly to the mechanism shown in the $\check{1}$  concave structures \autoref{fig:evoconcave434_1Lzoom}, a collective movement of atomic columns leads to the nucleation of a new axis in a more central position.

\begin{table}[h!]
\small
\centering

\begin{tabular}{cclccc}

\hline
size & best & $T_{melt}$ (K) & 500 K & 550 K & 600 K \\

\hline 

\textbf{262}& Dh &625-645 & 100\% & 100\% & --   \\

\textbf{382}& Dh & $\sim 690$ & --   & 100\% & 100\% \\

\hline

\end{tabular}
\caption{Final percentage of Dh after 1\ \textmu s as a function of temperature and cluster size, for $\hat{2}$ structures derived from Dh348 and Dh520 decahedra.}
\label{tab:evolution_convex2}

\end{table}

\section{Discussion and conclusions}
Our results demonstrate that the stability and evolution of fivefold-twinned gold nanoparticles are governed by a geometry-controlled competition between surface-driven diffusion and strain-driven detwinning. The fivefold axis, which can be interpreted as a wedge disclination embedded in a finite crystal, responds sensitively to both surface curvature and its depth beneath the surface.

Two dominant behaviors emerge. In concave geometries, surface diffusion progressively fills the re-entrant region, lowering surface energy and promoting either re-centering of the original disclination or nucleation of a new, more centrally located fivefold axis. Even in cases where the decahedral motif is not the global minimum for a given size, concavity provides a kinetic and thermodynamic bias toward restoring fivefold symmetry. In this regime, geometry effectively stabilizes the topological defect.

In contrast, convex structures with the fivefold axis located directly beneath a single surface layer undergo rapid detwinning within the first nanoseconds of evolution. Here, glide of the superficial atomic layer removes two twin planes and eliminates the fivefold rotational arrangement, relieving elastic strain at the expense of surface compactness. The resulting single-twinned configuration may subsequently stabilize, fully detwin into an FCC particle, or, more rarely, retwin through the nucleation of a new disclination.

A key finding is the strong sensitivity of defect stability to axis depth. When the fivefold axis lies beneath two atomic layers, detwinning is completely suppressed within the timescales explored, and the system consistently evolves toward re-centering. The additional atomic layer acts as a mechanical confinement, inhibiting surface glide and effectively increasing the kinetic barrier for disclination annihilation. Thus, a change of only one atomic layer in axis depth qualitatively alters the evolutionary pathway.

A critical question is whether the re-centering mechanisms observed here persist as nanoparticle size increases toward experimentally relevant scales. In the size range investigated (200–600 atoms), decahedral gold structures remain thermodynamically competitive because the reduction in surface energy compensates for the elastic strain associated with the disclination axis. However, as particle size increases, the volumetric strain energy (scaling approximately with particle volume) grows faster than the surface energy contribution. In larger particles, the internal pressure at the core may promote the emission of partial dislocations or the formation of stacking faults to relieve strain, even in otherwise symmetric structures.

For convex configurations, two atomic layers may no longer be sufficient to stabilize the axis at larger scales. The increased internal strain would likely reduce the relative energy barrier for detwinning, making a peripheral axis more prone to annihilation before surface diffusion can re-center it. In contrast, the concavity-mediated re-centering mechanism appears more robust with increasing size. Experimental observations of oriented attachment in larger gold nanoparticles support this picture, showing that concave junctions facilitate the emergence and centering of five-fold twins by providing a favorable pathway for atomic rearrangement.

Overall, across all simulated systems, with the exception of convex particles featuring a single superficial layer, the tendency toward axis re-centering or retwinning remains dominant. This structural preference is consistent with experimental reports that small gold decahedra typically exhibit a centered five-fold axis. Our results reproduce this tendency and clarify the atomistic mechanisms that drive the disclination back toward the particle core. More broadly, the study highlights how surface geometry and defect depth govern the competition between strain relief and symmetry restoration in multi-twinned nanoparticles.

\section*{CRediT authorship contribution statement}
\textbf{Silvia Fasce}: Validation, Formal analysis, Investigation, Data Curation, Writing - Original Draft.
\textbf{Diana Nelli}: Writing - Review \& Editing, Supervision.
\textbf{Luca Benzi}: Methodology, Validation.
\textbf{Georg Daniel Förster}: Methodology.
\textbf{Riccardo Ferrando}: Conceptualization, Writing - Review \& Editing, Supervision, Funding acquisition.

\section*{Declaration of competing interest}
The authors declare that they have no known competing financial interests or personal relationships that could have appeared to influence the work reported in this paper.

\section*{Acknowledgements}
The authors acknowledge financial supports under the National Recovery and Resilience Plan (NRRP), Italy, Mission 4, Component 2, Investment 1.1, Call for tender No. 104 published on 2.2.2022 by the Italian Ministry of University and Research (MUR), funded by the European Union – NextGenerationEU – Project Title PINENUT – CUP D53D23002340006 - Grant Assignment Decree No. 957 adopted on 30/06/2023 by the Italian Ministry of University and Research (MUR). The authors acknowledge networking support from the IRN Nanoalloys of CNRS.

\section*{Data Availability}
The spatial coordinates of global minima and initial structures for molecular dynamics simulations and the complete trajectories of the evolution sequences shown in Figures \ref{fig:348_conv1_md4_600K}, \ref{fig:evoconcave434_1Lzoom}, \ref{fig:evo_concave_0Lnewaxis}, \ref{fig:348_conv0_md2_zoom}, \ref{fig:evo_348conv1L_detwinning_press}, \ref{fig:evo_348conv1L_twinning} and \ref{fig:evoconv348_2L} are available on GitHub \cite{gitHub_repo}. Additional data generated or analysed during this study are available upon request from the corresponding author.

\printbibliography

\end{document}